# CFD CHARACTERIZATION OF STREET CANYON HEATING BY SOLAR RADIATION ON BUILDING WALLS

*Beatrice Pulvirenti[1], Silvana Di Sabatino[2]*

[1]Department of Industrial Engineering, University of Bologna, Italy
[2]Department of Physics and Astronomy, University of Bologna, Italy

**Abstract**: Heat exchange processes between building walls and external air within street canyons is an important topic in air quality modelling considering that the thermal fields largely affect local flow dynamics and pollutant concentration distribution. Despite the abundance of numerical studies, many questions still remain unanswered, ultimately limiting the inclusion of heat effects in mesocale atmospheric models. Of interest in this study is the assessment of radiative processes at building walls in terms of emissivity (i.e. is the ratio of the thermal radiation from a surface to the radiation from an ideal black surface at the same temperature), shape factors (i.e. the proportion of the radiation which leaves the surface of a building and strikes toward the surface of another building) and their relation to the Richardson number. The study is approached via the computational fluid dynamics code OpenFOAM, with large eddy simulations (LES) extension to model turbulence within the canyon and Boussinesq approach to model heat transfer. Consequences on pollutant concentrations are also analysed. A model for setting the thermal boundary conditions on the building surfaces is developed, based on a database that gives the correlations between the emissivity of a surface and the heat flux emitted by this surface, under certain weather conditions, as a function of the latitude and longitude of the surface, the orientation, the time of the day. As a novel result an exponential law between the transfer coefficient and Ri (as well as with the emissivity) is proposed as a function of the canyon aspect ratio.

**INTRODUCTION**
Due to rapid urbanization and industrialization, streets have become one of the most common areas where both pedestrian and traffic density are usually high. Until now, interest in the study of flow and pollutant dispersion in urban street canyons has rapidly increased. To achieve a better understanding of street canyon flow and pollutant dispersion and to provide useful guidelines to improve air quality in densely built cities, a large number of investigations have been carried out through field and wind tunnel experiments as well as numerical model simulations (Ahmad, K. et al., 2005). These investigations allowed the identification of many factors affecting flow and dispersion in street canyons, such as the ambient wind speed and direction, street canyon aspect ratio, building area density and so on. Thermal effects are additional effects that need to be addressed in detail. In urban areas, those effects are due to solar radiation reaching building walls and ground surfaces that in return heat up air in the vicinity. The role of buoyancy forces within street canyon is particularly relevant under calm wind conditions, when downward inertial forces are compensated by an upward flow due to the buoyancy. Numerical studies of buoyant flows are less numerous in the literature with respect to those using isothermal conditions. Most numerical investigations are limited to two-dimensional simulations using steady Reynolds Averaged Navier–Stokes equations (RANS) models (e.g. Xie, et al., 2007) or unsteady RANS models (Sini et al., 1996; Kim and Baik, 2001). Two-dimensional studies usually do not account for the highly three-dimensional flow fields occurring in real urban street canyons of finite length. Using a three-dimensional, steady RANS model, Tsai, et al. (2005) have examined thermal effects of heated building walls on flow and pollutant dispersion in a street canyon of aspect ratios H/W=0.8 and L/W=3 (where H is the height, W the width and L the length of the street canyon). They showed that the vortex line that connects the centres of the cross sectional vortices meanders in the street canyon. Only few studies report on three-dimensional numerical simulations of thermal effects. Coupling thermal and dispersion charactreistics in street canyons are rare in the literature due to the high computational cost. In this paper, we investigate the impact of ground and wall heating on flow and pollutant dispersion in a street canyon by means of the computational fluid dynamics (CFD) code OpenFOAM by employing a one-equation subgrid-scale (SGS) LES model. Three street canyon aspect ratios have been considered, H/W=0.5, 1 and 2. To avoid unrealistic three-dimensional effects, depending on L/H, a ratio L/H=20 has been chosen following the results of several premiliminary tests.

**MODELLING SETUP**
In this problem, Navier-Stokes equations for fluid motion, together with Fourier equation for temperature distribution of fluid flow have been considered. The Boussinesq approximation has been assumed. In direct numerical simulation (DNS) the Navier-Stokes equations are solved as such using fine spatial and time resolution. In LES, the resolution is coarser and Navier-Stokes equations (NS) are solved for the spatially filtered variables pressure and velocity, with additional subgrid scale (SGS) terms appearing as source terms in the equations. The LES equations take the form NS(u,p)=$\tau_{sgs}$, where $\tau_{sgs}$ requires modeling. Here, we choose the one-equation eddy-viscosity SGS approach by Yoshizawa (Yoshizawa, 1993) where an additional transport equation for the SGS turbulent kinetic energy is solved:

$$\frac{\partial k_{sgs}}{\partial t} + \frac{\partial}{\partial x_j}(u_j k_{sgs}) = \frac{\partial}{\partial x_j}\left[(v + v_{sgs})\frac{\partial k_{sgs}}{\partial x_j}\right] + P_{k_{sgs}} - C_\epsilon \frac{k_{sgs}^{3/2}}{\Delta}$$

(1)

where $P_{ksgs}$ is the resolved rate of strain tensor and $V$ is the volume of the computational cell. The filtered LES equations can then be closed by setting $\tau_{sgs}$. The spatial discretization utilizes the second order accurate finite volume method. The computational domain is a parallelepiped with dimensions 320 m in the x direction, parallel to the flow direction, 310 m in the y direction and 80 m in the vertical z direction. The buildings are two parallelepipeds with length (L) equal to 200 m and width (W) equal to 10 m. The height H of the canyon varies from 10 to 20 m, in order to give aspect ratios in the range 0.5<W/H<2. The choice of the building dimensions has been the result of a wide number of preliminary tests devoted to understand the dimensionality of the problem. We have found that for L/H<15 the velocity field is strongly 3D, as discussed in the next section. In this paper, L/H=20 for all the aspect ratio is considered.
The dimensions of outer domain gives an appropriate mesh size for the required flow detail and run time. The computational domain has been built using structured elements with a finer resolution close to the ground and the walls within the canyon. A finer refinement with respect to the isothermal simulations has been needed in order to capture thermal gradients near the walls. Several tests have been performed to verify grid size independence with increasing mesh numbers. The final number of the computational cells used is about 1 million cells for all the cases. The smallest dimension of the elements, in the region near the heated walls, is 0.25 m in the direction normal to the wall and 0.5 m in the other directions.
The dimensionless number governing natural convection in this problem is the Richardson number Ri=Gr/Re$^2$ . At the inlet section, ambient air temperature at ground level has been assumed to 27°C. Cases with and without solar radiation at different locations, such as heating at ground level, the leeward and windward sides of buildings are analysed (as shown in Figure 1) by imposing different emissivity of the surfaces, that translates in different temperatures of the walls and different incident heat fluxes.

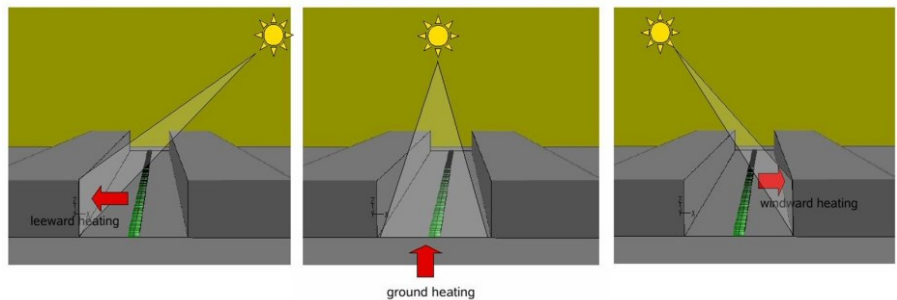

**Figure 1**. Representation of different wall heating from the sun.

The effect of solar radiation to the walls with different values of the emissivity has been quantified as an incident heat flux to the surfaces. This translates in different Richardson numbers. For example, walls with emissivity e=0.95 gives a value of Ri=3.2.

The inlet boundary condition is a logarithmic boundary layer velocity profile similar to that used in our previous works (Di Sabatino, S. et al. 2007), where the friction velocity $u_*$ has been modified in order to obtain a given average wind speed u(z) at the height H. Inlet turbulent kinetic energy and dissipation rate profiles are specified as Di Sabatino, S. et al. (2007). Symmetry boundary conditions are specified on the top and lateral sides of the computational domain.

**RESULTS**

A number of preliminary tests has been performed, in order to investigate the dimensionality of the problem. We have found that, for L/H<15 the velocity field is strongly 3D. To avoid these 3D effects, in this paper, L/H=20 has been used, for all the aspect ratios W/H considered. With this length to height ratio L/H=20, the following cases have been considered, H/W=0.5;1;2.

Figure 2 shows path lines coloured by z-component of velocity, on a vertical plane in the middle of the canyon. These results refer to the case Ri=3.27. For leeward heating a weak dependence on the aspect ratio is observed, as the vortex within the canyon is enhanced by the buoyancy. For ground heating we can observe that the vortex is weaker as the aspect ratio increases; a higher and higher z-component of velocity in the central region could break the vortex for higher aspect ratios. For windward heating, the classical counter-rotating vortex is observed for W/H=2. Many authors have found this configuration for the case of windward heating. The vortex becomes dominant as the aspect ratio W/H decreases and for W/H=0.5 the 'cold clockwise' vortex is suppressed by the counter-rotating vortex, that occupies the whole region of the canyon. The scalar fluxes from the street canyon are important quantities for the transfer processes between the urban canopy layer and the overlying atmosphere. Mesoscale models or models of urban energy balance require the parameterization of these transfer processes.

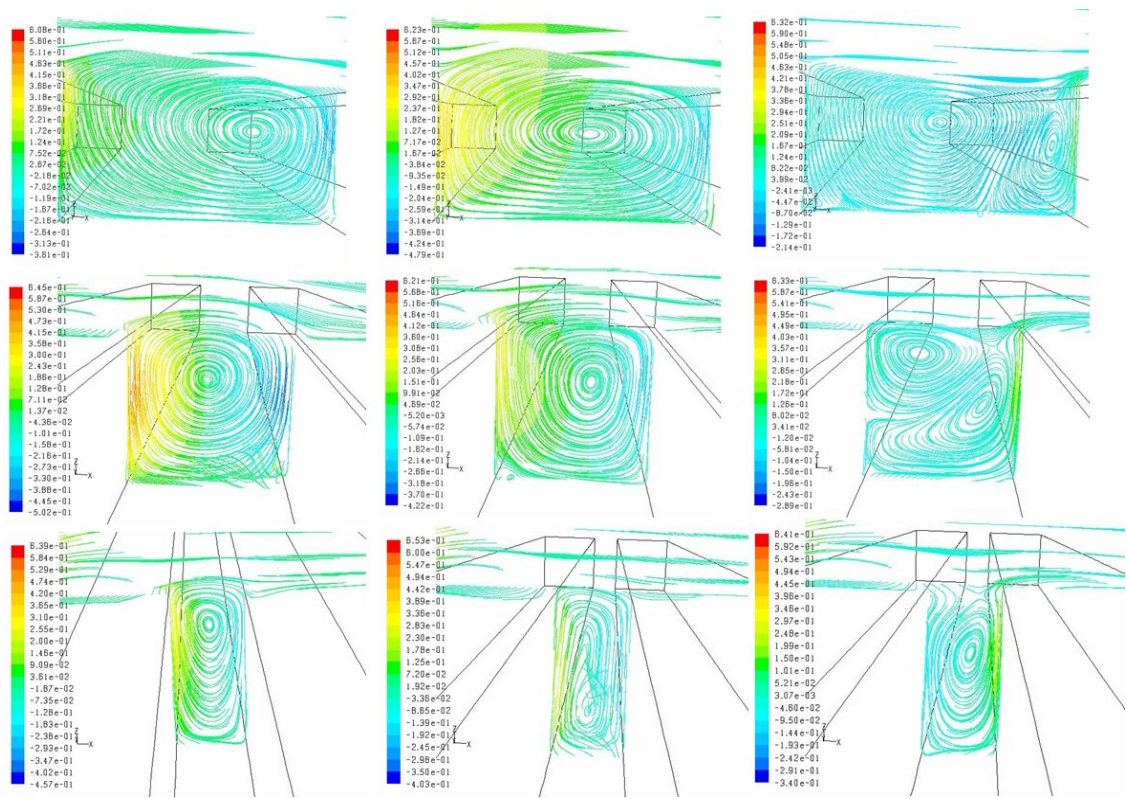

**Figure 2**. Path lines for Ri=3.27. Aspect ratios: W/H=2 (top), W/H=1 (middle) and W/H=0.5 (bottom). Leeward heating (left), ground heating (middle) and winward heating (right).

Therefore, many numerical studies have been performed to quantify these processes (Li et al. 2016). If the spatio-temporal average of the scalar flux at the roof level is F and the "strength" of the source is Cs (this can be either the temperature for heat flux or the concentration for pollutant flux), and the ambient (background) "strength" of the scalar is assumed to be zero, the transfer coefficient is then defined as:

$$\phi = F/(U\,Cs) \tag{1}$$

where F is the heat flux and Cs is the surface temperature; F is often referred to as the Stanton number (St). Note that here the transfer coefficient is defined for either the air within the street canyon or for a specific facet of the street canyon. F and $\phi$ can be split into parts due to advection and turbulence:

$$F^{tot} = F^{adv} + F^{turb} \quad \text{and} \quad \phi^{tot} = \phi^{adv} + \phi^{turb} \tag{2}$$

Previous studies have shown that the transfer coefficients and exchange velocities depend on the urban geometry, locations of sources, and atmospheric stability.

Figure 3 shows the transfer coefficients of the passive pollutant from a line source at the street level, for the case H/W=1. The advective parts of the transfer coefficients are close to 0 for all Ri. The turbulent parts are of the order of $10^{-4}$ for pollutant transfer coefficient and of $10^{-3}$ for heat transfer coefficient. Both coefficients scale with Ri as the total. An exponential law between the coefficient and Ri (as well as with the emissivity) can be obtained, for each value of the canyon aspect ratio.

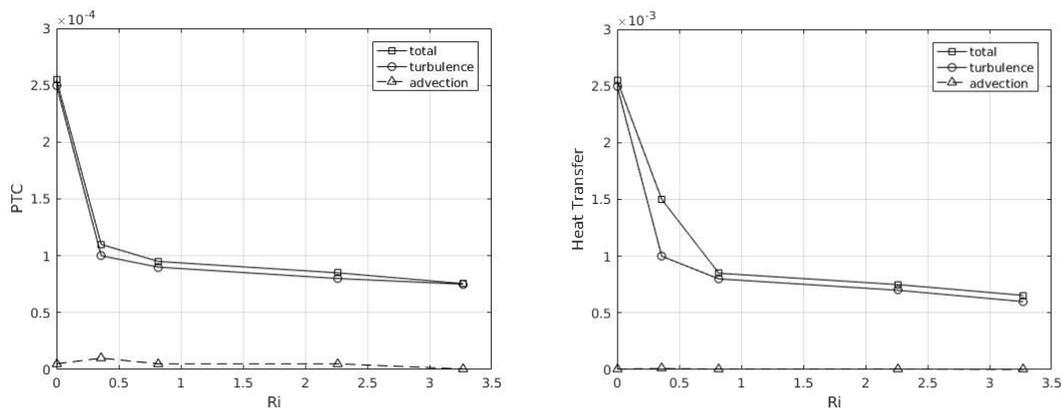

**Figure 2.** Pollutant transfer coefficient (left) and heat transfer coefficient (right).

**CONCLUSIONS**

In the present study, a LES code has been employed to study the effect of thermal radiation on the dispersion characteristics within the street-canyon. The effect of solar radiation to walls with different values of the emissivity has been quantified as an incident heat flux to the surfaces. The results show that, with leeward heating, the canyon vortex is enhanced and its centre is shifted against the wind direction causing lower concentrations to occur. In the case of windward heating, convection leads to the introduction of colder air from the bottom of the canyon forming a second vortex close to the windward wall. In this case larger pollutant concentrations near the windward wall and within the whole street canyon are found. LES results show that the advective parts of the transfer coefficients are close to 0 for all Ri, while the turbulent parts are of the order of $10^{-3}$ and scale with Ri as the total. An exponential law between the pollutant transfer coefficient (PTC) and Ri (as well as with the emissivity) can be obtained, for each value of the canyon aspect ratio. This novel result can be further exploited for an easier inclusion of thermal effects in street canyons within mesoscale atmospheric models.